~r paper
\documentstyle[12pt]{article}
\begin{document}
\title{
Supersymmetry Breaking Threshold Corrections\\
in the $SU(4)\times SU(2)_L\times SU(2)_R$ Model}
\vspace{1cm}
\author{
O. Korakianitis$^{a,b}$\\
and\\
N. D. Tracas$^a$
\vspace{1cm}\\
a: Physics Department\\
National Technical University\\
GR-157 80 Zografou\\
Athens, Greece
\vspace{.7cm}\\
b: Physics Department\\
Royal Holloway and Bedford New College\\
University of London\\
Egham, Surrey TW20 OEX, UK
}
\date{}
\maketitle
\begin{abstract}
\noindent
We evaluate the SUSY and top threshold effects in the context of the
MSSM and the string derived model based on
SU(4)$\times$SU(2)$_L\times$SU(2)$_R$. In both cases we run the two
loop RGEs and determine the lower bounds of the supersymmetric particle
masses, dictated by the experimentally accepted regions of the values
of the low energy parameters.
\end{abstract}
\thispagestyle{empty}
\vspace*{6.0cm}

\begin{flushleft}
NTUA 41/93 \\
May 1993
\end{flushleft}
\vfill

\newpage
\setcounter{page}{1}
{\bf I. Introduction}
\vspace{.5cm}

The success of the Standard Model (SM) is beyond question. It provides
understanding of the strong and the electroweak interactions, while
(within the experimental and theoretical errors) it has
succeeded to pass all experimental tests, so far. Nevertheless, the
plethora of free parameters, in conjuction with the question of
unification at high energies, remains an unsolved problem. As far as
the second issue is concerned, Supersymmetric (SUSY) models show,
in general, a better attitude providing coupling constant unification
at scales of the order $10^{16-17}$GeV, while at the same time reproduce
the experimental values of the low energy parameters.

The non-observation of SUSY particles forces us to admit the
existence of an energy region, above $M_Z$, where the (non-SUSY) SM is
effective. The present accuracy of measuring the low energy parameters
($\alpha^{-1}$, $s^2$ and $\alpha_3$) permits us to check the limits of
successfulness of the SUSY models through the threshold effects
\cite{wein,hall,ross}
of the SUSY particles to the running of the coupling constants.

In this letter, after a quick overview of the threshold corrections,
we concentrate on the Minimal Supersymmetric Standard Model (MSSM) and
the SUSY thresholds, and we find their effects on the low energy
parameters under several scenaria of SUSY masses. Since the region of
the SUSY particle masses extends as low as $M_Z$, we cannot ignore the
threshold effects due to the top quark. Finally, we incorporate the
SUSY (and top) threshold effects to a successful Grand Unified model,
namely the one based on the $SU(4)\times SU(2)_L\times SU(2)_R$
symmetry and check its ability to reproduce the low energy
parameters.

\vspace{1cm}
{\bf II. The threshold corrections}
\vspace{.5cm}

The Renormalozation Group Equations (RGE's) govern the running of the
gauge couplings through the corresponding $\beta$-functions of the
groups on which our theory is based. These $\beta$-functions are
determined by the light particle content of the model. If at some
energy scale our symmetry breaks to a smaller one, some of the
particles of the initial symmetry eventualy become massive and do not
contribute to the $\beta$-function below that scale. The threshold
corrections take into account the contribution of these massive states
to the running of the gauge couplings, since they could appear as
virtual particles even below the symmetry breaking scale. The effect
is the same as if we had subtracted each particle contribution to the
$\beta$-function(s) at the energy scale equal to its mass. Following
the formalism or Refs.
\cite{wein,hall},
in the vicinity of the symmetry breaking scale, the coupling constants
of the unbroken $\alpha _G$ and of the broken $\alpha _{G_k}$ regions
are related by
\[
\frac{1}{\alpha _{G_k}(\mu)} \equiv
\frac{1}{\alpha_G(\mu)}-\delta(\alpha^{-1}_{G_k}) =
\frac{1}{\alpha_G(\mu)}-4\pi\lambda _k(\mu)
\]
It is easy to see that these corrections are of the same order with
the two-loop solution of the RGE's for the gauge couplings. The
``matching'' function $\lambda$ is given by the general formula
\cite{lang,carena}
\[
\lambda(\mu)= 2\sum _i b_i
ln\frac{M_i}{\mu}+C
\]
where $b_i$ is the contribution to the $\beta$-function of the
particle i with mass $M_i$. The constant term depends on the
renormalization scheme. In the $\overline{DR}$ scheme $C=0$, while in
the $\overline{MS}$ scheme it is given by
\[C=-\frac{2}{21}b_i'\]
where $b_i'$ is the contribution to the
$\beta$-function of the gauge bosons (plus corresponding ghosts and
would be goldstone bosons) acquiring mass. This term appears in the
$\overline{MS}$ scheme from the product of the $1/\epsilon$-term of
the momentum integral with the $\epsilon$-term of the $\gamma$-matrix
algebra. This latter is missing in the $\overline{DR}$ scheme. The
connection between the two schemes comes through the conversion
factor
\[
\frac{1}{\alpha_{\overline{DR}}}=\frac{1}{\alpha_{\overline{MS}}}-
\frac{C_2(G)}{12\pi}
\]
where $C_2(G)$ is the quadratic Casimir
operator of the group $G$. By changing from one scheme to another and
back, the difference of the two conversion factors gives the constant
term of the $\overline{MS}$ scheme.

An equivalent formula for the matching function is the following
\cite{hall}
\[
\lambda(\mu)=\frac{1}{48\pi ^2}
             \left(Tr[t_{iV}^2]-21 Tr[t_{iV}^2ln\frac{M_V}{\mu}]+
                                C_FTr[t_{iF}^2ln\frac{M_F}{\mu}]+
                                C_STr[t_{iS}^2ln\frac{M_S}{\mu}]
             \right)\]
where $V$, $F$, and $S$ stand for vector boson, fermion and scalar,
while $t$ is the group generator in the appropriate representation.
The index $i$ runs over the particles that become massive. The factor
for the fermions takes the value $C_F=8$ for Dirac and $C_F=4$ for
two-component spinors or Majorana fermions (the latter is applicable
when L and R components are treated separately). For the scalars the
values are $C_S=1$ for complex representations and $C_S=2$ for real
ones. For a SUSY theory the above formula takes the form
\[
\lambda(\mu)=\frac{1}{48\pi ^2}
             \left(Tr[t_{iV}^2]-13 Tr[t_{iV}^2ln\frac{M_V}{\mu}]+
                              6 Tr[t_{iC}^2ln\frac{M_C}{\mu}]
             \right)\]
where $V$ and $C$ stand for the vector and chiral multiplets. Note the
unchanged constant term that comes from the spin-1 momentum integral.
As before, in the $\overline{DR}$ scheme this constant term is missing.

\vspace{1cm}
{\bf III. SUSY and Top Thresholds}

\vspace{.5cm}
We now turn on the threshold corrections corresponding to the breaking
of supersymmetry, but we focus only on the MSSM breaking down to the
(non SUSY) SM. The three matching functions
corresponding to the groups $SU(3)$, $SU(2)$ and $U(1)$ are
\begin{eqnarray}
\lambda_3(\mu)&=&
\frac{1}{48\pi^2}\left(12ln\frac{M_{\tilde{g}}}{\mu}+
                       12ln\frac{M_{\tilde{q}}}{\mu}-3\right)\\
\lambda_2(\mu)&=&
\frac{1}{48\pi^2}\left(9ln\frac{M_{\tilde{q}}}{\mu}+
                       8ln\frac{M_{\tilde{W}}}{\mu}+
                       3ln\frac{M_{\tilde{l}}}{\mu}+
                       4ln\frac{M_{\tilde{H}}}{\mu}+
                        ln\frac{M_{H}}{\mu}-2\right)\\
\lambda_1(\mu)&=&
\frac{1}{48\pi^2}\left(\frac{33}{5}ln\frac{M_{\tilde{q}}}{\mu}+
                       \frac{27}{5}ln\frac{M_{\tilde{l}}}{\mu}+
                       \frac{12}{5}ln\frac{M_{\tilde{H}}}{\mu}+
                       \frac{3}{5}ln\frac{M_{H}}{\mu}\right)
\end{eqnarray}
where the subscripts $\tilde{q}$, $\tilde{g}$, $\tilde{l}$,
$\tilde{W}$ and $\tilde{H}$ stand for
the squark, gluino, slepton, wino and higgsino, while $H$ stands for the
heavy Higgs doublet. In the above formulae, we have incorporated the
constant term coming from the conversion from $\overline{DR}$, above
SUSY breaking, to $\overline{MS}$ below.

Let us find the order of magnitude of the above matching functions
assuming several naive scenaria for the masses. In order to choose the
scale $\mu$ we ``admit'' that life is supersymmetric and run the gauge
couplings down to $M_Z$ using the MSSM $\beta$-functions. Then
$\mu=M_Z$. We take $\alpha_3=.12$, $sin^2\theta_W\equiv s^2=.232$ and
$\alpha=(127.9)^{-1}$.

\begin{itemize}
\item
Suppose that all masses are degenerate, of the order of 1TeV.
Then we get
\[4\pi\lambda_3(M_Z)\simeq 1.44,\hspace{1cm}
  4\pi\lambda_2(M_Z)\simeq 1.54,\hspace{1cm}
  4\pi\lambda_1(M_Z)\simeq 0.96\]
By writting the SUSY threshold corrected parameters in the form
\begin{equation}
\alpha_3+\delta^{thr}(\alpha_3),\hspace{1cm}
  s^2-\delta^{thr}(s^2),          \hspace{1cm}
  \alpha^{-1}-\delta^{thr}(\alpha^{-1})
\end{equation}
we get, at $M_Z$,
\[\delta^{thr}(\alpha_3)\simeq 0.02,\hspace{1cm}
  \delta^{thr}(s^2)\simeq 0.006,\hspace{1cm}
  \delta^{thr}(\alpha^{-1})\simeq 3.1\]
\item
The second idea is to assume a common mass $M_c$ for the coloured
particles and a common mass $M_{weak}$ for the rest of them. The
results are summarized in Table I, for selected values of $M_c$ and
$M_{weak}$, and in Figs. (1a) and (1b), where we draw contours of
constant $\delta^{thr}$ in the space $(M_c,M_{weak})$.
\end{itemize}

Naively it seems strange that the heavier the SUSY particle masses the
stronger their effects. But remember that we run the MSSM
$\beta$-functions down to $M_Z$. Therefore the heavier the particles
the bigger the energy region where the non SUSY $\beta$-functions
should have been used. Or, alternatively, if your
favorite SUSY (down to $M_Z$) model gives you low energy parameter
values outside the experimentally accepted regions, choosing the
appropriate SUSY particle masses, could possibly rescue your model
when you include the threshold effects.

Finally, we present the $\delta^{thr}$ for the case where the SUSY
particle masses are given in terms of a universal gaugino mass
$m_{1/2}$ and a universal scalar mass $m_0$, at $M_X$, according to
the (one loop) equations
\cite{barin}

\[m_{\tilde{g}}=\frac{\alpha_3(m_{\tilde{g}})}{\alpha_U}m_{1/2},
\hspace{1cm}
m_{\tilde{W}}=\frac{\alpha_2(m_{\tilde{W}})}{\alpha_U}m_{1/2}\]
and
\[m^2_{\tilde{q}}=m^2_0+7m^2_{1/2},
\hspace{.5cm}
m^2_{\tilde{l}}=m^2_0+0.3m^2_{1/2}\]
where we have assumed a common squark mass and a common slepton mass
(and also that $tan\beta =<\bar{v}>/<v>=1$).
Using the known values of
$\alpha_3$ and $\alpha_2$ at $M_Z$, we express the mass of
$m_{\tilde{W}}$, $m_{\tilde{q}}$ and $m_{\tilde{l}}$ as a function of
$m_{\tilde{g}}$ and $m_0$. In Figs. (2a) and (2b) we show contours of
constant mass for $m_{\tilde{W}}$, $m_{\tilde{q}}$ and
$m_{\tilde{l}}$ in the ($m_{\tilde{g}},m_0$) space. In the same space,
in Figs. (3a), (3b) and (3c) we show contours of constant
$\delta^{thr}$'s. We have chosen $m_{\tilde{H}}=m_H=100$ GeV and
$\alpha_U=(25)^{-1}$. It is evident that in the region where the mass
of the sparticle is smaller than $M_Z$, the corresponding term in the
matching function is missing.

As a general comment we can state that the threshold effects push the
coupling $\alpha_3$ to higher values, while the opposite happens for
$s^2$ and $\alpha^{-1}$ (see our definitions in Eqs(4)). The simple
reason is the missing of the negative term in the $\lambda$-functions
which corresponds to the spin-1 particle, while the negative term from
the conversion is too small to reverse the situation (even if
$m_{\tilde{q}}=m_{\tilde{g}}\ge 130$GeV, while all other sparticles are
below $M_Z$, all $\lambda$-functions are positive).

A heavy top can also give threshold effects which should be taken into
account. It is easy to evaluate the corresponding contribution to
$\alpha _3$ and $\alpha$. They are given by
\[\delta^{top}(\alpha_3^{-1})=\frac{1}{3\pi}ln\frac{m_t}{M_Z},
\hspace{1cm}
  \delta^{top}(\alpha^{-1})=\frac{8}{9\pi}ln\frac{m_t}{M_Z}\]
As far as the correction to $s^2$ is concerned, following Ref.
\cite{lang}
we use the formula
\[\delta^{top}(s^2)=1.05\times 10^{-7}{\rm GeV}^{-2}
                                    (m_t^2-M_Z^2)\]
The $\delta^{top}$ corrections are defined in an analogous way to
Eq.(4). In Table II we show the corrections $\delta
^{top}(\alpha^{-1})$, $\delta^{top}(s^2)$ and $\delta ^{top}(a_3)$,
for different values of $m_{top}$.

In Table III we show inputs and results from running the RGE's (of the
MSSM) from an initial energy $E_X$ and an initial $\alpha_X$ down to
$M_Z$, where all threshold effects (SUSY and top) are taken into
account. Finally, in Fig. 4 we show the region (shaded), in the
space ($m_{\tilde{g}},m_0$), which allows experimentally accepted
values for the low energy parameters. We vary the mass of the top
in the range $120{\rm GeV}<m_t<180$GeV and the unification constant in
the range $0.039<\alpha_X<0.041$. The mass of the higgs (and higgsino)
is chosen to be 150 GeV. This allowed region can give some crude lower
bounds for the SUSY particles
\[
m_{\tilde{g}}>250{\rm GeV}\hspace{.5cm}
m_{\tilde{W}}>80{\rm GeV},\hspace{.5cm}
m_{\tilde{l}}>75{\rm GeV},\hspace{.5cm}
m_{\tilde{q}}>220{\rm GeV}
\]
We have to note that although the allowed range of $m_t$ was the one
mentioned above, the values suggested by the threshold effects were
mostly in the upper half of that range.

\vspace{1cm}
{\bf IV. Threshold Effects in the
SU(4)$\times$SU(2)$_L\times$SU(2)$_R$ Model}

\vspace{.5cm}
In this last section we are going to evaluate the threshold
corrections to a string derived model based on
SU(4)$\times$SU(2)$_L\times$SU(2)$_R$. The symmetry group, derived
from the free fermionic formulation, is
SU(4)$\times$O(4)$\times$U(1)$^4\times$\{SU(8)$\times$U(1)$^\prime$\}
$_{Hidden}$
\cite{al}.
For more information on the spectrum and properties of the model see
Ref.
\cite{al,lt1}.
We quickly review the different scales appearing in the running of
the RGEs:
\begin{itemize}
\item
$M_U$, where $\alpha_4=\alpha_{2L}=\alpha_{2R}\equiv\alpha_U$
\item
$M_A$, where one of the U(1)'s, which is anomalous, breaks and a number
of fields acquire masses through some singlet fields. Between $M_U$
and $M_A$ we assume the full string content of the model. Both $M_U$
and $M_A$ can be fairly good approximated by the simple expressions
\[M_U\sim 1.7 \sqrt{4\pi\alpha_U}\cdot 10^{18}{\rm GeV}\hspace{1cm}
  M_A\sim 7.8 \sqrt{4\pi\alpha_U}\cdot 10^{17}{\rm GeV}\]
\item
$M_X$, where the group SU(4)$\times$SU(2)$_L\times$SU(2)$_R$ breaks
down to the MSSM and the relations among the couplings are
\begin{equation}
\alpha_3=\alpha_4, \hspace {1cm}
\alpha_2=\alpha_{2L}, \hspace{1cm}
\frac{1}{\alpha_1}=\frac{1}{\frac{5}{3}\alpha_{2R}}+
                   \frac{1}{\frac{5}{2}\alpha_4}
\end{equation}
and finally
\item
$M_I$, where we assume that below that scale we only have the standard
model content. Between $M_X$ and $M_I$ some exotic remnants could
survive.
\end{itemize}

 We give the breaking
pattern at $M_X$ for the different multiplets (the quantum numbers on
the left correspond to SU(4)$\times$SU(2)$_L\times$SU(2)$_R$ while on
the right to SU(3)$\times$SU(2)$_L$ $\times$U(1))
\begin{eqnarray*}
n_H(^(\overline{4}^),1,2)&\rightarrow &
    n^\prime_{31}(^(\overline{3}^),1,\pm 2/3)+
    n_3(^(\overline{3}^),1,\pm 1/3)+
    n_s(1,1,\pm 1)+n^\prime _s(1,1,0)\nonumber \\
n_4(^(\overline{4}^),1,1)&\rightarrow &
    n^\prime _3(^(\overline{3}^),1,\pm 1/6)+
    n^\prime(1,1,\pm 1/2)\nonumber \\
n_{22}(1,2,2)&\rightarrow &
    n_2(1,2,\pm 1/2)\nonumber \\
n_L(1,2,1)&\rightarrow&
    n^\prime _L(1,2,0)\nonumber \\
n_R(1,1,2)&\rightarrow &
    n^\prime (1,1,\pm 1/2)\\
n_6(6,1,1)&\rightarrow &n_3(^(\overline{3}^),1,\pm 1/3)
\end{eqnarray*}
The number of generations is always 3. The full string content is
\[
n_g=3,\hspace{.5cm} n_L=n_R=10,\hspace{.5cm}
n_6=n_{22}=n_H=2n_4=4\]

In a previous work
\cite{ltthr}
the threshold effects of the fields becoming massive at $M_X$, were
taken into account, assuming a degenerate mass, of the order of $M_X$.
The MSSM was effective down to $M_Z$. It was found that, for $M_X$
in the region $(3\cdot 10^{15}-10^{16})$GeV, the low energy parameters
stay inside the experimental limits for a wide range of particle
content between $M_A$ and $M_X$ and with a remnant
($n_3^{\prime}=n^{\prime}=2$) between $M_X$ and $M_I$. What values of
$\alpha_3$, $s^2$ and $\alpha^{-1}$ should we expect at $M_Z$ from the
RGEs if the subsequent addition of SUSY and top thesholds drive us
inside the experimental limits? $\alpha_3$ should stay low while $s^2$
and $\alpha^{-1}$ should get high. This could be achieved by lowering
the $\alpha_U$. But this does not seem to be enough to keep from
$\alpha_3$ running high at $M_Z$. A way out is to lower $M_X$ or $M_I$
or to keep more remnants (e.g. $n_3^{\prime}=4$) between $M_X$ and
$M_I$. Another way to control $s^2$ is through the difference
$n_L-n_R$ (the tendency of the low energy parameters to all these
changes is explained in Ref.
\cite{ltthr})

In Fig.5 we show, in the parameter space ($m_{\tilde{g}},m_0$), the
allowed region for $m_t=200$GeV. The inputs are the following
\[
\alpha_U=.053,\hspace{.5cm}
M_X=10^{15}{\rm GeV},\hspace{.5cm}
M_I=(10^{12}-10^{13}){\rm GeV}
\]
\[
n_L=n_R=4,\hspace{.5cm}
n_H=n_{22}=n_4=\frac{n_6}{2}=2,\hspace{.5cm}
n_3^{\prime}=n^{\prime}=2,\hspace{.5cm}
n_{31}^{\prime}=n_L^{\prime}=0
\]
In the GUT threshold we have used a degenerate mass of the order
$M_X/\sqrt 2$. We have taken into account that the three low energy
couplings are not any more unified at $M_X$ and the masses of
$m_{\tilde{W}}$, $m_{\tilde{q}}$ and $m_{\tilde{l}}$ are given by
the approximate formulae
\[
m_{\tilde{W}}\sim .33 m_{\tilde{g}},\hspace{.5cm}
m_{\tilde{q}}^2=m_0^2+1.2m_{\tilde{g}}^2,\hspace{.5cm}
m_{\tilde{l}}^2=m_0^2+.05m_{\tilde{g}}^2
\]
The higgs and higgsino masses are taken to be 100GeV.
The region comes out from a tricky interplay between all three
thresholds. It corresponds, roughly, to the mass regions
\[
m_{\tilde{W}}\le 125{\rm GeV},\hspace{.5cm}
m_{\tilde{l}}\ge 115{\rm GeV},\hspace{.5cm}
240{\rm GeV}\le m_{\tilde{q}}\le 460{\rm GeV}
\]

We manage to run the RGE's for higher $M_X=10^{16}$GeV but the price
to pay was a lower $M_I=10^{11}$GeV. We had also to change the
($n_L,n_R$) pair to (6,0). Only a small region in the left of the
phase space, namely $m_{\tilde{g}}\le 150$GeV and $100$GeV$\le m_0\le
170$GeV, and $120$GeV$\le m_t\le 140$GeV is allowed.

\vspace{1cm}
{\bf IV. Conclusions}

\vspace{.5cm}
We have evaluated the SUSY (and top) thresholds which should be taken
into account when the MSSM RGE's are being run down to $M_Z$. For a
simple GUT model with unification energy of the order of $10^{16}$GeV
and $a_U\sim .04$, we can get the experimentally allowed values of the
low energy parameters for
 $120{\rm GeV}<m_t<180$GeV, while the masses of the sparticles should
obey the following bounds: $m_{\tilde{g}}<$250GeV,
 $m_{\tilde{W}}>80$GeV,
$m_{\tilde{l}}>75$GeV and $m_{\tilde{q}}>220$GeV.
For the specific string derived model the situation gets complicated
but a general remark is that the GUT and the intermediate scale should be
lowered with respect to the case where no thresholds were taken into
account. The problem depends mostly on $\alpha_3$, since it has the
tendency to run high, at $M_Z$, while the threshold effects make the
situation worse. Therefore, in the case of this specific model, the
task is to keep $\alpha_3$ low.

\vspace{1.5cm}
We would like to thank G.K. Leontaris, S.D.P. Vlassopulos, M.
Carena and C. Wagner for helpful discussions.

This work is partially supported by a C.E.C. Science Program
SCI-CT91-0729.

\newpage

\begin{thebibliography}{9}

\bibitem{wein}S. Weinberg, Phys. Lett. 91B(1980)51;
\bibitem{hall}L. Hall, Nucl. Phys. B187(1981)75.
\bibitem{ross}G.G. Ross and R.G. Roberts, Nucl. Phys. B377(1992)571.
\bibitem{lang} P. Langacker and N. Polonsky, Univ. of Pennsylvania
preprint, UPR-0513T (1992).
\bibitem{carena}M. Carena, S. Pokorski and C.E.M. Wagner, MPI
preprint, MPI-Ph/93-10 (1993).
\bibitem{barin}R. Barbieri, S. Ferrara, L. Maiani, F. Palumbo and C.A.
Savoy, Phys. Lett. 115B(1982)212;

K. Inoue, A. Kakuto, H. Komatsu and S. Takeshita, Progr. Theor. Phys.
68(1982)297 and 71(1984)413.
\bibitem{al}I. Antoniadis and G.K. Leontaris, Phys. Lett.
216B(1989)333;

I. Antoniadis, G. K. Leontaris and J. Rizos, Phys. Lett.
245B(1990)161.
\bibitem{lt1}G.K. Leontaris and N.D. Tracas, Phys. Lett.
260B(1991)339 and  Z. Phys. C56(1992)479.
\bibitem{ltthr}G.K. Leontaris and N.D. Tracas, Phys. Lett.
291B(1992)44.
\vfill

\newpage

{\bf Table and Figure Captions}

\vspace{1cm}
{\bf Table I}: SUSY thresholds for $\alpha^{-1}$, $s^2$ and $\alpha_3$
for the degenerate case of 1 TeV for all masses and for two other
cases in the $M_c$, $M_{weak}$ scenario.

\vspace{.5cm}
{\bf Table II}: Top thresholds for different $m_t$

\vspace{.5cm}
{\bf Table III}: The low energy parameter values, with SUSY and top
thresholds included, in the context of the MSSM for three selected
values of $M_X$, $\alpha_X$ and sparticle masses. We show two cases
for the higgs and higgsino masses and for several $m_t$.

\vspace{.5cm}
{\bf Fig.1}: Contours of constant $\delta^{thr}(s^2)$ (a) and constant
$\delta^{thr}(\alpha^{-1})$ and $\delta^{thr}(\alpha_3)$ (b) in the
$M_c$ , $M_{weak}$ scenario. The $\delta^{thr}(\alpha_3)$ contours
corresponds to the vertical lines since they depend only on $M_c$.

\vspace{.5cm}
{\bf Fig.2}: Contours of constant mass for the sleptons and winos (a)
and for the squarks (b) in the ($m_{\tilde{g}}$, $m_0$)
space. The wino contours correspond to the vertical lines since they
depend only on $m_{\tilde{g}}$.

\vspace{.5cm}
{\bf Fig.3}: Contours of constant $\delta^{thr}(\alpha^{-1})$ (a),
$\delta^{thr}(s^2)$ (b) and $\delta^{thr}(\alpha_3)$ (c) in the
($m_{\tilde{g}}$, $m_0$) space, in the context of the MSSM.

\vspace{.5cm}
{\bf Fig.4}: Allowed (shaded) region in the ($m_{\tilde{g}}$,
$m_0$) space, for 120GeV$<m_t<$180GeV, .039$<\alpha_X<$.041,
in the context of the MSSM.
Unification energy $E_X=10^{16}$GeV and $m_{\tilde{H}}=m_H=150$GeV.

\vspace{.5cm}
{\bf Fig.5}: Allowed region in the ($m_{\tilde{g}}$, $m_0$)
space, for $m_t=200$GeV, in the context of the
SU(4)$\times$SU(2)$_L\times$SU(2)$_R$ model. The breaking to the SM
group is performed at $M_X=10^{15}$GeV and $m_{\tilde{H}}=m_H=100$GeV.

\vfill
\newpage
{\bf Table I}
\vspace{.5cm}

\begin{tabular}{||c||c|c|c||}
\hline\hline
&$\delta^{thr}(\alpha^{-1})$&$\delta^{thr}(s^2)$&$\delta^{thr}(\alpha_3)$\\
                                                    \hline\hline
$M_c=M_{weak}=1TeV$& 3.1    & 0.006   & 0.02    \\ \hline
$M_c=256GeV$       &        &         &         \\
                   & 1.2    & 0.002   & 0.01    \\
$M_{weak}=200GeV$  &        &         &         \\ \hline
$M_c=200GeV$       &        &         &         \\
                   & 0.8    & 0.001   & 0.006   \\
$M_{weak}=150GeV$  &         &         &        \\ \hline\hline
\end{tabular}

\vspace{1.5cm}
{\bf Table II}

\vspace{.5cm}
\begin{tabular}{||c||c|c|c||}
\hline\hline
$m_{top}$(GeV)&$\delta^{top}(\alpha ^{-1})$ &
           $\delta^{top}(s^2)$ & $\delta^{top}(\alpha_3)$\\
\hline\hline
100 & 0.03 & 0.0002  &  10$^{-4}$\\
\hline
120 & 0.08 & 0.0006  &  4$\cdot 10^{-4}$\\
\hline
140 & 0.12 & 0.0012  &  7$\cdot 10^{-4}$\\
\hline
160 & 0.16 & 0.0018  &  9$\cdot 10^{-4}$\\
\hline
180 & 0.19 & 0.0025  &  10$^{-3}$\\
\hline
200 & 0.22 & 0.0033  &  1.2$\cdot 10^{-3}$\\
\hline\hline
\end{tabular}

\vspace{1.5cm}
{\bf Table III}

\setlength{\unitlength}{1cm}
\begin{picture}(12,8)(0,-1)
\put(0,5)
{
  \begin{picture}(3.2,3.2)
    \begin{tabular}{|l|}
    \hline
    $E_X=2\cdot 10^{16}$GeV\\
    $\alpha_X=0.0418$\\
    $M_{\tilde{g}}=300$GeV\\
    $M_{\tilde{W}}=90$GeV\\
    $M_{\tilde{q}}=330$GeV\\
    $M_{\tilde{l}}=162$GeV\\
    \hline
    \end{tabular}
  \end{picture}
}
\put(4,5.1)
{
  \begin{picture}(6.5,2.8)
    \begin{tabular}{|c|c|c|c|}\hline\hline
    \multicolumn{4}{|l|}{$M_H=M_{\tilde{H}}=100$GeV}\\ \hline\hline
    $m_t$(GeV) & $\alpha^{-1}$ &  $s^2$  &  $\alpha_3$ \\
    \hline
    160        & 128.0         & 0.2317   &    0.125\\
    \hline
    180        & 128.0         & 0.2310   &    0.125\\
    \hline
    200        & 128.0         & 0.2302  &    0.125\\
    \hline
    \end{tabular}
  \end{picture}
}
\put(4,1.2)
{
  \begin{picture}(6.5,2.3)
     \begin{tabular}{|c|c|c|c|}\hline\hline
    \multicolumn{4}{|l|}{$M_H=M_{\tilde{H}}=150$GeV}\\ \hline\hline
     $m_t$(GeV) & $\alpha^{-1}$ &  $s^2$  &  $\alpha_3$ \\
     \hline
     120        & 128.0         & 0.2327  &    0.125\\
     \hline
     140        & 128.0         & 0.2322  &    0.125\\
     \hline
     160        & 127.9         & 0.2315  &    0.125\\
     \hline
     180        & 127.9         & 0.2308  &    0.125\\
     \hline
     200        & 127.9         & 0.2300  &    0.125\\
     \hline
     \end{tabular}
  \end{picture}
}
\end{picture}

\begin{picture}(12,8)(0,-1.5)
\put(0,5)
{
  \begin{picture}(3.2,3.2)
    \begin{tabular}{|l|}
    \hline
    $E_X=2\cdot 10^{16}$GeV\\
    $\alpha_X=0.0418$\\
    $M_{\tilde{g}}=400$GeV\\
    $M_{\tilde{W}}=122$GeV\\
    $M_{\tilde{q}}=425$GeV\\
    $M_{\tilde{l}}=170$GeV\\
    \hline
    \end{tabular}
  \end{picture}
}
\put(4,4.8)
{
  \begin{picture}(6.5,2.8)
    \begin{tabular}{|c|c|c|c|}\hline\hline
    \multicolumn{4}{|l|}{$M_H=M_{\tilde{H}}=100$GeV}\\ \hline\hline
    $m_t$(GeV) & $\alpha^{-1}$ &  $s^2$  &  $\alpha_3$ \\
    \hline
    120        & 127.9         & 0.2323   &    0.127\\
    \hline
    140        & 127.9         & 0.2318   &    0.127\\
    \hline
    160        & 127.8         & 0.2312   &    0.127\\
    \hline
    180        & 127.8         & 0.2304   &    0.127\\
    \hline
    \end{tabular}
  \end{picture}
}
\put(4,1.4)
{
  \begin{picture}(6.5,2.3)
     \begin{tabular}{|c|c|c|c|}\hline\hline
    \multicolumn{4}{|l|}{$M_H=M_{\tilde{H}}=150$GeV}\\ \hline\hline
     $m_t$(GeV) & $\alpha^{-1}$ &  $s^2$  &  $\alpha_3$ \\
     \hline
     120        & 127.8         & 0.2321  &    0.127\\
     \hline
     140        & 127.8         & 0.2316  &    0.127\\
     \hline
     \end{tabular}
  \end{picture}
}
\end{picture}

\begin{picture}(12,8)(0,-1)
\put(0,5)
{
  \begin{picture}(3.2,3.2)
    \begin{tabular}{|l|}
    \hline
    $E_X=10^{16}$GeV\\
    $\alpha_X=0.0407$\\
    $M_{\tilde{g}}=400$GeV\\
    $M_{\tilde{W}}=122$GeV\\
    $M_{\tilde{q}}=417$GeV\\
    $M_{\tilde{l}}=170$GeV\\
    \hline
    \end{tabular}
  \end{picture}
}
\put(4,5)
{
  \begin{picture}(6.5,2.6)
    \begin{tabular}{|c|c|c|c|}\hline\hline
    \multicolumn{4}{|l|}{$M_H=M_{\tilde{H}}=100$GeV}\\ \hline\hline
    $m_t$(GeV) & $\alpha^{-1}$ &  $s^2$  &  $\alpha_3$ \\
    \hline
    200        & 128.0         & 0.2328   &    0.114\\
    \hline
    220        & 128.0         & 0.2319   &    0.114\\
    \hline
    240        & 128.0         & 0.2310   &    0.114\\
    \hline
    \end{tabular}
  \end{picture}
}
\put(4,1.7)
{
  \begin{picture}(6.5,2.3)
     \begin{tabular}{|c|c|c|c|}\hline\hline
    \multicolumn{4}{|l|}{$M_H=M_{\tilde{H}}=150$GeV}\\ \hline\hline
     $m_t$(GeV) & $\alpha^{-1}$ &  $s^2$  &  $\alpha_3$ \\
     \hline
     200        & 127.9         & 0.2326  &    0.114\\
     \hline
     220        & 127.9         & 0.2317  &    0.114\\
     \hline
     240        &  127.9        & 0.2308   &   0.114\\
     \hline
     \end{tabular}
  \end{picture}
}
\end{picture}

\end{document}